\newcommand{\dur}{\langle S(t)\rangle}
\newcommand{\ncov}{\langle n_{\rm cov}(t)\rangle}

\documentstyle[multicol,aps,psfig]{revtex}
\begin{document}

\draft

\title{Universal persistence exponents in an extremally driven system}

\author{D. A. Head\footnote{Current address:
Division of Physics and Astronomy,
Vrije Universiteit, De Boelelaan 1081,
NL-1081 HV Amselveen, NL}}

\address{Department of Physics and Astronomy,
JCMB King's Buildings, University of Edinburgh,
Edinburgh EH9 3JZ, UK}

\date{\today}

\maketitle

\begin{abstract}
The local persistence $R(t)$, defined as the proportion
of the system still in its initial state at time $t$,
is measured for the Bak--Sneppen model.
For 1 and 2 dimensions, it is found that the decay
of $R(t)$ depends on one of two classes of initial configuration.
For a subcritical initial state, $R(t)\sim t^{-\theta}$,
where the persistence exponent $\theta$ can be expressed
in terms of a known universal exponent.
Hence $\theta$ is universal.
Conversely, starting from a supercritical state,
$R(t)$ decays by the anomalous form
$1-R(t)\sim t^{\tau_{\rm ALL}}$ until a finite time~$t_{0}$, where
$\tau_{\rm ALL}$ is also a known exponent.
Finally, for the high dimensional model $R(t)$ decays
exponentially with a non--universal decay constant.
\end{abstract}


\maketitle
\begin{multicols}{2}
\narrowtext


Extremally
driven systems represent an important
class of non--equilibrium models,
with applications including
sheared granular media, biological macroevolution
and interface
depinning~\cite{ShearedGM,Rev,BS,WeakMem}.
Their defining characteristic is that they are driven
in the vicinity of an {\em extremal} quantity,
often the minimum or maximum of
an inhomogeneous scalar field.
Such deterministic dynamics can be naturally realised in
the limit of vanishing (thermal) noise~\cite{EPJB,Vergeles,SnepT}.
Moreover, many extremal models are also
`critical' in an analogous sense to a
continuous phase transition in equilibrium
systems~\cite{SOCBook}.

However, the temporal evolution of these systems
is in many cases not fully understood.
This is clearly demonstrated by recent work on one
of the simplest, and therefore most studied,
extremally driven systems, namely the
Bak--Sneppen model~\cite{BS}.
This has been shown to {\em age} in a
similar manner to glassy systems~\cite{BSAge,JPhysA},
suggesting that it does not reach a steady state
in the timeframe of current simulation methods.
This is in contrast to the usual claim that stationarity
is reached
{\em `after an extensive transient'}~\cite{BS}.
It now appears that this contradiction
arises because most studies only measure
functions of a single time variable,
such as critical exponents {\em etc.},
that can appear to become constant after a finite time
even for a non--stationary system;
a true test of stationarity (such as
the aging result mentioned above)
require the measurement of two--time
correlation functions~\cite{Kob,GlassRev}.

The aim of this paper is to investigate one aspect of
the temporal behaviour of the Bak--Sneppen
that has not yet been studied, in the hope that this
will further elucidate the time evolution
of extremally driven systems in general.
The quantity that we consider is
the local {\em persistence}~$R(t)$,
defined as the proportion of the system at a time
$t$ that has not yet changed from its initial state.
This quantity has been the focus of considerable
attention in recent years, at least in part because it is often
found to decay algebraically with a non--trivial
{\em persistence exponent}~$\theta$,
$R(t)\sim{\rm (const)}\,t^{-\theta}$,
that (for equilibrium systems)
cannot be related to either the static or
dynamic exponents~\cite{Derrida,Bray,Majumdar,Haye}.
The universality of $\theta$ remains an open subject.
In many cases it is not universal; that is, it may depend
on the lattice coordination number 
or the precise choice
of interaction term~\cite{Haye,Sudhir}.
However, it has been found to be universal in directed
percolation~\cite{Haye}.
We find this to also be the case for the Bak--Sneppen model,
and confirm our numerical finding by
deriving an expression for $\theta$ in terms of
the known universal exponent $\gamma$~\cite{Rev}.
We further find that, when starting from a supercritical initial
configuration, $R(t)$ decays in an anomalous,
non--algebraic manner.
Finally, $R(t)$ always decays exponentially
the infinite dimensional case.

\noindent{\em The model:} The Bak--Sneppen model
is defined as follows~\cite{Rev,BS}.
A scalar quantity $f_{i}\in[0,1]$ is
independently assigned to every site $i$
of a $d$--dimensional hypercubic lattice,
which contains a total of $N$ sites.
For every unit time~$t$, the smallest $f_{i}$ in the system
is found, and it and its nearest neighbours are assigned new
values, again drawn uniformly from $[0,1]$.
Other initial conditions and
interaction terms will be considered below.
To generate the random values of the $f_{i}$\,,
we used L'Ecuyer's 64 bit combined multiple recursive
pseudo--random number generator MRG63k3a,
which has a quoted period of $2^{377}$~\cite{Ecuyer}.
The typical number of operations required to find the global
minimum was reduced from
${\cal{O}}(N)$ to ${\cal{O}}(\ln N)$
by using a binary search tree,
which also constrains the system size to be a power of
2~\cite{Grassberger}.

\begin{figure}
\centerline{\psfig{file=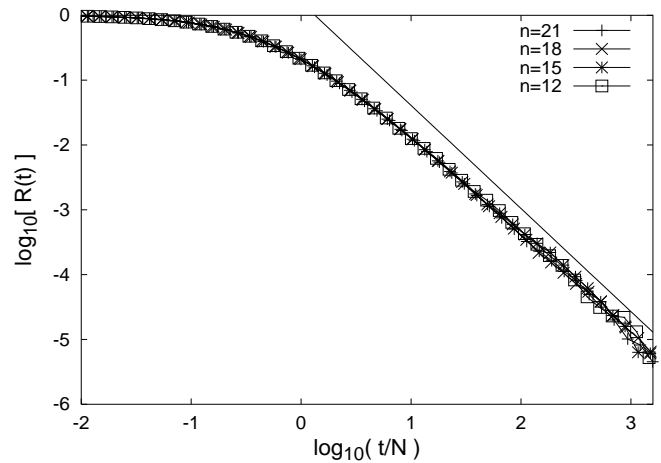,width=9cm}}
\vspace{0.3cm}
\caption{The persistence $R(t)$ against the rescaled time
variable $t/N$ for different system sizes
$N=2^{n}$, with $n=12$, 15, 18 and 21 as shown.
The thin solid line has a slope of -1.59.}
\label{f:1d_size}
\end{figure}

Typical results for a periodic 1D lattice are presented
in Fig.~\ref{f:1d_size}.
It can clearly be seen that the curves of $R(t)$ for
different system sizes collapse after rescaling the
time scale by $N$, for all $N>2^{10}\approx10^{3}$.
For the largest system $N=2^{21}$,
the simulation time was long and only
one run was performed; however, for smaller systems
$R(t)$ was averaged over 
10-100 different initial configurations.
Also plotted in the figure is the theoretical prediction
$R(t)\propto t^{-1.59}$, which will be derived below.
The numerical $R(t)$ appear to be converging towards
this prediction at late times, but there is still a slight curvature
to the lines (on log--log axes) even for the largest $t$
we were able to simulate, making direct
measurement of the exponent difficult.

\begin{figure}
\centerline{\psfig{file=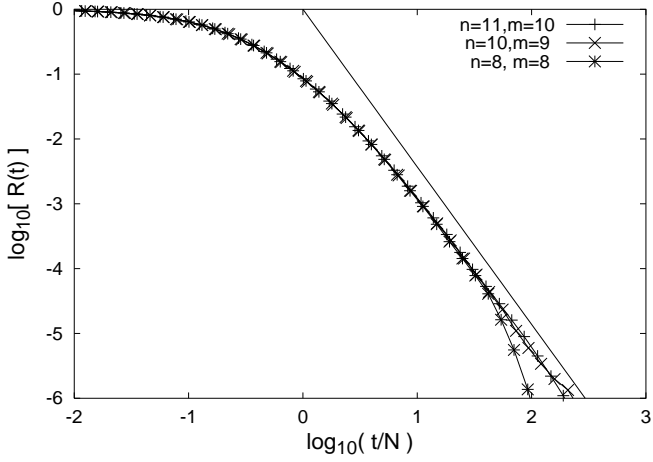,width=9cm}}
\vspace{0.3cm}
\caption{The persistence $R(t)$ for two dimensional
lattices of sizes $N=2^{m}\times 2^{n}$.
The solid straight line has a slope of -2.43.
The data was averaged over at least 10 separate runs.}
\label{f:2d_size}
\end{figure}

A similar picture emerges in two dimensions,
as shown in Fig.~\ref{f:2d_size}.
The theoretical prediction in this case $R(t)\propto t^{-2.43}$
(again, see below for details), which is consistent with
the simulation data, although as before there is still
some slight curvature to $R(t)$ for the largest $t$
available to our current computing resources.

\begin{figure}
\centerline{\psfig{file=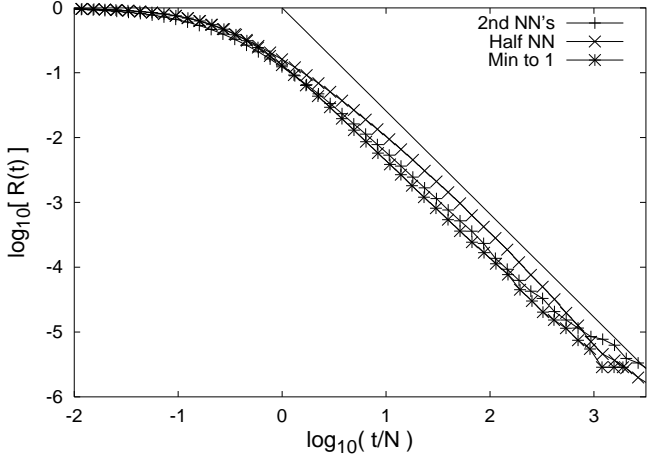,width=9cm}}
\vspace{0.3cm}
\caption{$R(t)$ for one dimensional systems of size
$L=2^{21}$ with different interaction rules, to determine
the universality of the persistence exponent.
See the main text for explanation of the terms
``2nd NN'', ``Half NN'' and ``Min to 1.''
The solid line has a slope of -1.59.}
\label{f:1d_univ}
\end{figure}

The universality of the exponent was tested by
altering the microscopic details of the interactions.
Three alternative rules were considered~:
updating the second nearest neighbours as well as
the nearest neighbours (``2nd NN's'');
changing the value of the nearest neighbours from
$f_{i}$ to $\frac{1}{2}(f_{i}+r)$, where the annealed random
variable $r$ is uniformly distributed on [0,1]
(``Half NN'');
and setting the minimum value to 1 but
changing the nearest neighbours as normal (``Min to 1'').
In all cases $\theta$ was found to be approximately
the same.
This has been tested in both one and two dimensions;
the 1D case is shown in Fig.~\ref{f:1d_univ}.

\noindent{\em Theory:} We shall now explain how
the persistence exponent can be theoretically derived.
This argument utilises known results for the Bak--Sneppen
model, which for reasons of space will not be repeated
here;
the reader is instead referred to \cite{Rev,BS,Maslov}
for their justification.
It is known that the location of the minimum or `active' site
initially jumps around the lattice in an uncorrelated manner,
but as the system evolves the lattice variables
$f_{i}$ become spatially correlated, and the active site
tends to remain
in a localised region of the lattice for a finite amount
of time before jumping to another, uncorrelated region
of the lattice.
Each such period of localised activity is known as
an `avalanche.'

Let the mean duration of an avalanche
at time $t$ be denoted by $\dur$, and the mean number of sites
changed at least once during the avalanche by $\ncov$,
where the angled brackets `$\langle\ldots\rangle$'
refer to averages over different initial configurations.
By taking a suitably large system size~$N$,
these quantities will increase by an arbitrarily small
amount between avalanches, allowing the continuum
limit to be taken.
It is known that both quantities increase algebraically
in time~\cite{Rev},

\begin{eqnarray}
\dur&\sim&A\left(\frac{t}{N}\right)^{\gamma/(\gamma-1)}\quad,
\label{e:dur}
\\
\ncov&\sim&B\left(\frac{t}{N}\right)^{1/(\gamma-1)}\quad.
\label{e:ncov}
\end{eqnarray}

\noindent{}Here, $\gamma$ is a universal exponent that depends
on the lattice dimensionality and the symmetries of the
interactions.
We assume $\gamma>1$;
The case of $\gamma=1$, realised in high dimensions,
will be discussed later.

By defintion, $R(t)=1$ at $t=0$ and decreases
by an amount $\delta R=1/N$ whenever
a lattice site changes its value for the first time.
The average rate at which sites are changed by
the avalanches is $\ncov/\dur$; however,
only a fraction $R(t)$ of sites have not already
changed from their initial state.
Recalling that the active site jumps to a random part
of the system inbetween avalanches,
then $R(t)$ obeys the following equation

\begin{equation}
\frac{{\rm d}R(t)}{{\rm d}t}
=
-
\frac{\ncov R(t)}{\dur N}\quad.
\label{e:R}
\end{equation}

\noindent{}Subsituting (\ref{e:dur}) and (\ref{e:ncov}) into 
(\ref{e:R}) gives

\begin{equation}
R(t)\sim {\rm (const)}\,t^{-\theta}
\end{equation}

\noindent{}with the persistence exponent $\theta=B/A$.

Since $A$ and $B$ are non--universal constant prefactors
to the algebraic growth of $\dur$ and $\ncov$ respectively,
one would expect that their ratio, and therefore $\theta$,
is also non--universal; however, a peculiar feature
of this model is that $B/A$ {\em is} universal.
This can be seen by combining together two exact
equations, referred to as the {\em gap} and {\em gamma}
equations in \cite{Rev}, which gives the following
equation relating $\dur$ and $\ncov$,

\begin{equation}
\frac{{\rm d}\dur}{{\rm d}t}
=
\frac{\ncov}{N}\quad.
\label{e:exact}
\end{equation}

\noindent{}Substituting (\ref{e:dur}) and (\ref{e:ncov}) 
into (\ref{e:exact}) gives
the persistence exponent to be

\begin{equation}
\theta
=
\frac{B}{A}
=
\frac{\gamma}{\gamma-1}\quad.
\end{equation}

\begin{figure}
\centerline{\psfig{file=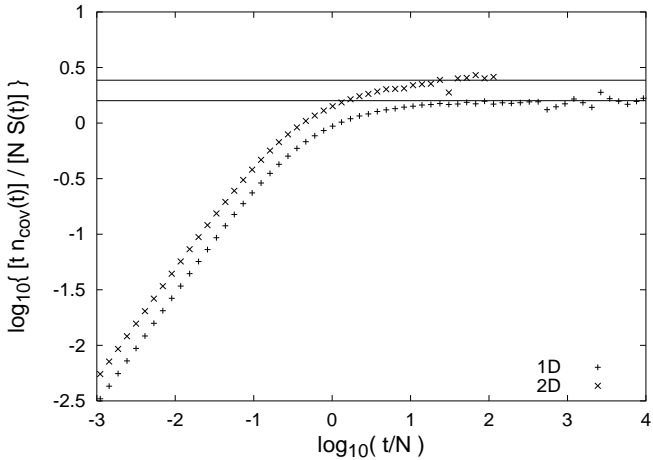,width=9cm}}
\vspace{0.3cm}
\caption{The quantity $t\ncov/(N\dur)$ against time
for 1 and 2 dimensional lattices of sizes $2^{21}$
and $2^{11}\times2^{10}$.
The argument in the main text predicts that
this will tend to a constant value that is the
persistence exponent.
For comparison, horizontal solid lines are plotted
for $\theta=1.59$ and $2.43$, which are the
exponents for 1D and 2D respectively.}
\label{f:SNcov}
\end{figure}

\noindent{}Since $\gamma$ is universal, then so is $\theta$.
In 1 and 2 dimensions, $\gamma=2.70$ and 1.70
respectively~\cite{Rev}, giving predicted values of
$\theta\approx 1.59$ and 2.43, which were the values
used in Figs.~\ref{f:1d_size}---\ref{f:1d_univ}.
As a further check on this analysis,
note that (\ref{e:dur}) and (\ref{e:ncov})
predict that the quantity
$t\ncov/(N\dur)$ will approach the constant
value $B/A=\theta$ at late times.
This is confirmed by the numerical results
presented in Fig.~\ref{f:SNcov}.

The preceding argument breaks down when
$\gamma=1$, which arises for dimensions above the upper
critical dimension~$d_{\rm c}$\,, for which the most recent
estimate is $d_{\rm c}=4$~\cite{ucd}.
In this case, $\dur$ and $\ncov$ both increase
exponentially in time rather than algebraically~\cite{Rev}.
By repeating the same steps as above,
it is straightforward to show that $R(t)$ should now
decay exponentially with a non--universal decay
constant $C$,

\begin{equation}
R(t)\sim{\rm (const)}\,{\rm e}^{-Ct}
\end{equation}

\begin{figure}
\centerline{\psfig{file=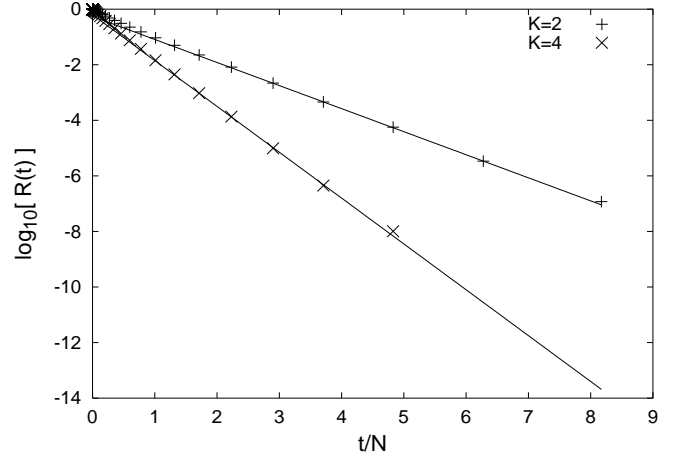,width=9cm}}
\vspace{0.3cm}
\caption{$R(t)$ for a random nearest neighbour model
in which the minimum and another $K-1$ sites chosen at
random from the lattice are assigned new values.
The system contained $N=2^{21}$ elements, and $R(t)$
was averaged over 100 runs.
The solid lines have slopes of -0.83 and -1.65.}
\label{f:RNN}
\end{figure}

\noindent{}This is confirmed by the numerical results
given in Fig.~\ref{f:RNN} for a system in which the minimum
and another $K-1$ randomly chosen elements are assigned
new values, which is the usual way of simulating an
infinite dimensional system for this model.
It is clear that the decay of $R(t)$ is exponential with a decay
constant that depends strongly on~$K$.

Finally, we consider the effects of initial conditions.
All of the preceding analysis and numerical results
has assumed an initial configuration in which every
$f_{i}$ is uniformly distributed over the range [0,1].
This can be generalised by choosing an initial
state in which the $f_{i}$ are uniformly distributed
over the range $[f_{0},1]$ and are spatially uncorrelated.
It is known that the Bak--Sneppen model has a
`critical value' $f_{\rm c}\in(0,1)$ such that the system is
{\em subcritical} if $f_{0}<f_{\rm c}$, and {\em supercritical}
if $f_{0}\geq f_{\rm c}$~\cite{Maslov}.
(The value of $f_{\rm c}$ is non--universal and depends
on the lattice structure and the microscopic interaction
term).
We have checked that any subcritical initial state gives
rise to the same persistence exponent in both one and
two dimensions, as predicted by our earlier argument.
However, for a supercritical initial configuration, the
system enters into a single, infinite avalanche, and all
of the preceding analysis fails.
In fact, $R(t)$ in this situation is simply related to the
growth of a single avalanche in time, which is already
known to increase algebraically in time with
a universal exponent $\tau_{\rm ALL}$~\cite{Rev}.
Thus $R(t)$ does not decay algebraically but rather
according to the expression

\begin{equation}
1-R(t)\sim {\rm (const)}\,t^{\tau_{\rm ALL}}
\label{e:seed}
\end{equation}

\begin{figure}
\centerline{\psfig{file=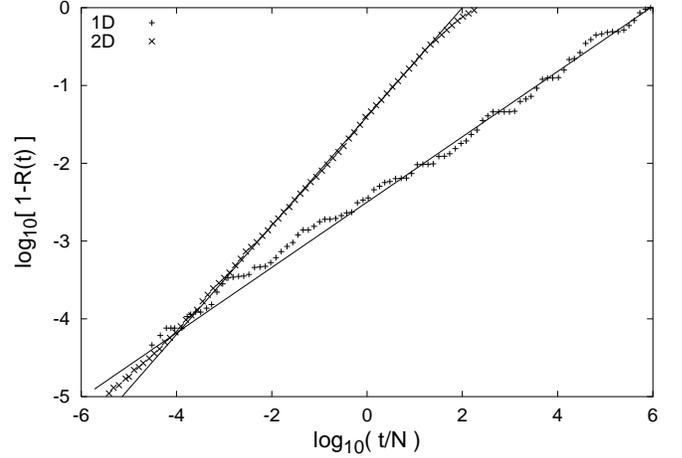,width=9cm}}
\vspace{0.3cm}
\caption{$1-R(t)$ versus $t/N$ for supercritical initial
conditions in which every element has a value $f_{i}=1$
except for a single `seed' site that has $f_{\rm seed}<1$.
The 1D data was obtained from a single run on
a $2^{16}$ lattice, and
the 2D data was averaged over 10 runs of a 
$2^{10}\times2^{10}$ lattice.
The solid lines give the known slopes of
$\tau_{\rm ALL}=0.42$ in 1D and 0.70 in 2D.
}
\label{f:seed}
\end{figure}

\noindent{}until a late time $t_{0}$ when the avalanche
`touches' itself through the periodic boundary
conditions.
Confirmation of this scaling is given in
Fig.~\ref{f:seed} for one and two dimensions.
In high dimensions, $R(t)$ decays exponentially
for both classes of initial state.

\noindent{\em Summary:} We have found that the persistence
$R(t)$ in the Bak--Sneppen model can decay in one
of three different ways.
In low dimensions,
it decays with a universal exponent $\theta$
when starting from a subcritical state, but by the anomalous
form $1-R(t)\propto t^{\tau_{\rm ALL}}$ when starting
from a supercritical state.
For high dimensions the decay is always exponential.
We also provided a simple theoretical prediction for
the value of $\theta=\gamma/(\gamma-1)$, which,
when combined with the numerical results, gives
strong confirmation of the universality of $\theta$.

It is not yet clear how this relates to the question of
the existence of a steady state of the model, although
it is interesting to note that $\theta>1$ always,
which is significantly higher than typical persistence
exponents, and means that the number of untouched
sites will rapidly decay to zero.
Thus the observed non--stationarity must be
due to temporal correlations within an avalanche,
rather than the spread of avalanches throughout the
system.

\noindent{\em Acknowledgements:} The author would
like to thank the hospitality of the Vrije Universiteit,
Amsterdam, where part of this work was carried out.
This work was funded by EPSRC (UK) grant no.
GR/M09674.



\end{multicols}

\end{document}